\documentclass[lettersize,journal]{IEEEtran}
\usepackage{amsmath,amsfonts}
\usepackage{algorithm}
\usepackage{array}
\usepackage[caption=false,font=normalsize,labelfont=sf,textfont=sf]{subfig}
\usepackage{textcomp}
\usepackage{stfloats}
\usepackage{url}
\usepackage{verbatim}
\usepackage{graphicx}
\usepackage{cite}

\usepackage{graphicx}

\usepackage{tabu}
\usepackage{wrapfig}
\usepackage[nodisplayskipstretch]{setspace}
\usepackage{geometry}
\usepackage{tcolorbox}
\usepackage{tikz}
 
\usepackage[noend]{algpseudocode}
\usepackage{multicol}
\usetikzlibrary{positioning, arrows.meta}
\DeclareMathOperator{\sinc}{sinc}

\usepackage{xspace}
\usetikzlibrary{arrows.meta,positioning,calc}
\usetikzlibrary{arrows.meta,positioning}
\usetikzlibrary{arrows.meta,positioning}
\tikzset{
	lc/.style   = {draw=yellow!70!black, fill=yellow!20, very thick,
		rounded corners, minimum width=16mm,
		minimum height=8mm, align=center},
	sc/.style   = {draw=cyan!70!black,   fill=cyan!15,   very thick, circle,
		minimum size=9mm, align=center},
	arr/.style  = {-{Latex[length=2.8mm]}, thick},
	sweep/.style= {->, very thick, dashed, red},
	annot/.style= {font=\footnotesize, text=gray!60!black}
}

\newcommand{\ph}{\Phi}
\newcommand{\dtau}{\Delta\tau}
\ifCLASSINFOpdf

\else

\fi

\begin{document}

\title{Intra-Pair Skew Propagation Graph (ISPG): An Analytical Model for Cascaded Channels}

\author{\IEEEauthorblockN{David Nozadze\IEEEauthorrefmark{1}\IEEEauthorrefmark{2},
Zurab Kiguradze\IEEEauthorrefmark{1}, Srinath Penugonda\IEEEauthorrefmark{1}, Sayed Ashraf Mamun\IEEEauthorrefmark{1}, Amendra Koul\IEEEauthorrefmark{1} and
Mike Sapozhnikov\IEEEauthorrefmark{1}}

\IEEEauthorblockA{\IEEEauthorrefmark{1} Cisco Systems Inc., San Jose, CA, USA }

\IEEEauthorblockA{\IEEEauthorrefmark{2} dnozadze@cisco.com}}
\markboth{IEEE TRANSACTIONS ON SIGNAL AND POWER INTEGRITY, VOL. num, 2026}%
{Shell \MakeLowercase{\textit{et al.}}: A Sample Article Using IEEEtran.cls for IEEE Journals}
\maketitle
 
\begin{abstract}
As data rates scale, intra-pair skew has become a critical bottleneck for high-speed differential signaling. Current analytical models are often limited, while 3D electromagnetic simulations are computationally intensive. This paper presents a comprehensive analytical framework for intra-pair skew in generic asymmetric coupled transmission lines, explicitly integrating skew into S-parameter formulations. We introduce the Intra-pair Skew Propagation Graph (ISPG), a novel graph-based methodology for calculating cumulative skew in complex, cascaded channels. The proposed framework is validated against both S-parameter simulations and empirical measurements of a 2m bulk twinax cable assembly, demonstrating excellent accuracy and robustness for high-speed interconnect design.
\end{abstract}
 \IEEEpeerreviewmaketitle
 

\section{Introduction}
The rapid evolution of high-speed digital systems has placed unprecedented demands on the design and modeling of interconnects. Differential signaling, implemented through coupled transmission lines, has become the de facto standard for achieving robust performance in multi-gigabit applications such as high-speed memory interfaces, backplanes, serial links, and SerDes channels. By suppressing common-mode noise and mitigating electromagnetic interference, differential signaling enables higher data rates while maintaining compliance with stringent electromagnetic compatibility (EMC) standards.

However, as data rates scale into tens and hundreds of gigabits per second, interconnect imperfections that were once negligible now emerge as critical limiting factors. Among these, intra-pair skew—the relative propagation delay between the two conductors of a differential pair—has become a dominant source of performance degradation. Skew may arise from fabrication asymmetries, dielectric inhomogeneity, or even subtle differences in trace routing. Even picoseconds of mismatch can impair timing margins, distort eye diagrams, and induce significant differential-to-common mode conversion, ultimately degrading bit error rates (BERs) in high-speed links.

Numerical simulation techniques such as 3D electromagnetic (EM) solvers can account for skew, but they do so at the cost of computational intensity and without yielding general closed-form insight. As a result, designers often rely on heuristic corrections or oversimplified assumptions when evaluating skew effects.

In recent years, several studies have investigated the impact of skew on system performance, often through measurement-based approaches or case-specific simulations \cite{2014_Tian_simp,2010_Miller_DC,2017_Nozadze_epeps,2007_Loyer_CT,2017_Baek_DC,Nalla_2017_EMC,Nozadze_2017_EMC,2021_Moon_spi, 2018_Koul_DC}. Nozadze et al. proposed SILD and FOM\_SILD as practical reciprocal metrics to evaluate intra-pair skew impact, showing strong correlation between FOM\_SILD and BER in 224 Gbps SerDes channels \cite{2026_Nozadze_SILD}. These prior works highlight the significance of intra-pair skew but do not provide a general analytical framework capable of describing its influence on coupled-line behavior in a form suitable for both theoretical understanding and practical application. In particular, there is still a need for closed-form expressions for complete channels comprising multiple loosely and strongly coupled segments. Such formulations should enable accurate predictions of system behavior and explicitly link intra-pair skew to S-parameters—fundamental to frequency-domain characterization and widely used in circuit simulators and design verification workflows.

This paper aims to address this gap by presenting a general derivation of intra-pair skew equations for generic coupled transmission lines, explicitly incorporating skew into the formulation of S-parameters. Unlike prior approaches that are limited to loosely coupled transmission lines, the proposed analysis captures skew effects in both loosely and strongly coupled transmission lines. The derived analytical equations provide clear insight into how skew modifies transmission and coupling characteristics, offering designers a practical analytical tool to predict skew-induced impairments in high-speed channels.

The paper is organized as follows. In Section II, we derive expressions for intra-pair skew in generic asymmetric coupled transmission lines, along with the corresponding single-ended and balanced S-parameters. Section III examines special cases and the behavior of these equations to provide better practical understanding, and presents correlations with simulation results. Section IV introduces a graph-based rules to predict intra-pair skew in complex channels consisting of multiple asymmetric, loosely coupled, and strongly coupled transmission lines. Section V validates the proposed skew calculation framework through empirical measurements of a 2m bulk twinax cable. The results demonstrate excellent agreement between analytical predictions and measured data across all configurations. Section VI provides the conclusions.
 
\section{Analytical Framework for Skew and S-Parameters} 

In this section, using transmission line theory, we will analytically derive the intra-pair skew equations and examine the impact of skew on both single-ended and differential S-parameters for a generic asymmetric coupled transmission line \cite{Mongia_1999}.

We begin by introducing the second-order matrix form of the Telegrapher’s equations (assuming the lossless case, i.e., resistance and conductance matrices are zero, $\mathbf{R} = \mathbf{G} = 0$), which describes the characteristics of two coupled asymmetric transmission lines as given by:
\begin{align}\label{eq:teleg}
	\frac{\partial^2 \bf{V}}{\partial x^2}+\omega^2{\bf{LC}\,\bf{V} }= 0\,,
\end{align}
where ${\bf{V}}=[V_1,V_2]$ are voltages at transmission lines and $\omega$ is an angular frequency. The inductance $\bf{L}$ and capacitance $\bf{C}$ matrices are
\begin{align}
\bf{L} =
\left[ {\begin{array}{cc}
		L_{11} & L_{12} \\
		L_{21} & L_{22} \\
\end{array} } \right], \;\;
\bf{C} =
\left[ {\begin{array}{cc}
		C_{11} & C_{12} \\
		C_{21} & C_{22} \\
\end{array} } \right].
\end{align}
The non-symmetrical matrices $\bf{L}$ and $\bf{C}$ models the asymmetry between P and N lines, which governs intra-pair skew.

The $c$- and $\pi$- mode phase velocities respectively are (for the details, see Appendix~\ref{App_A}).
\[
v_{c(\pi)} =\sqrt{\frac{2}{L_{11} C_{11} + L_{12} C_{21} + L_{21} C_{12} + L_{22} C_{22} \mp u}}.
\]
 \begin{figure}[!t]
	\includegraphics[width=3.5in]{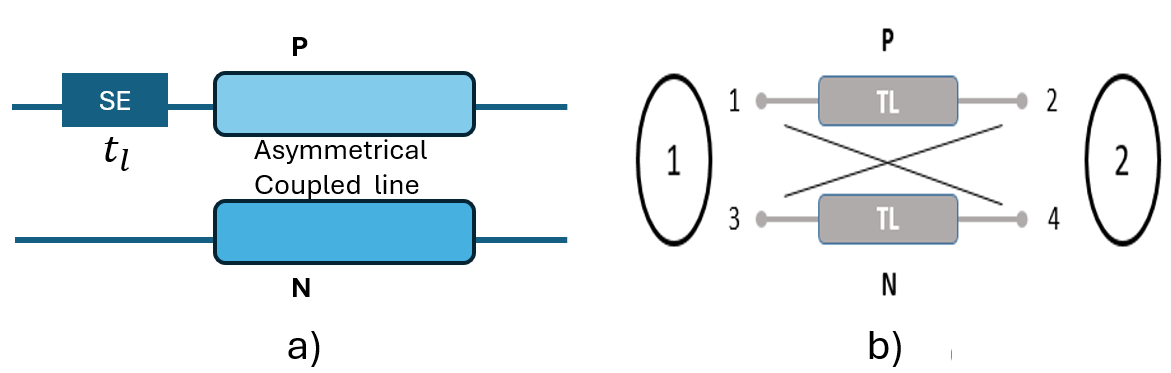}
	\caption{(a) Schematic of simulated channels: intra-pair skew is introduced by placing single-ended (SE) delay lines in front of the  asymmetrical coupled transmission line. The delay of the delayed lines $t_l$ is adjusted to vary the intra-pair skew. (b) Schematic of coupled transmission line: Ports labeled as 1 and 2 represent mixed-mode ports, while ports labeled as 1 through 4 correspond to single-ended ports. The P/N lines form a differential pair.}
	\label{FIG:one}
\end{figure}
The single-ended S-parameter matrix of an asymmetric coupled line, with an additional intra-pair skew $t_{l}$ introduced on one of the lines (P or N) as a frequency independent time delay, can be expressed as follows (Fig.~\ref{FIG:one} a): 
\begin{align}\label{S_param_00}
\left[ {\begin{array}{cc}
		S_{21}  & S_{41}  \\
		S_{23} & S_{43} \\
\end{array} } \right]=
\left[ {\begin{array}{cc}
				S_{21}^o e^{i 2\pi f t_{l}}  & S_{41}^o e^{i 2\pi f t_{l}}  \\
				S_{23}^o & S_{43}^o \\
		\end{array} } \right],
\end{align}
where
$$
S_{21}^o = \left(\cos\left(\pi f \Delta \tau \right)-i \frac{1-p}{1+p} \sin\left(\pi f\Delta \tau \right)\right) e^{-i \pi f \ell\left( v_c^{-1}+v_{\pi}^{-1}\right)},
$$
and 
$$
S_{43}^o = \left(\cos\left(\pi f\Delta \tau \right)+i \frac{1-p}{1+p} \sin\left(\pi f\Delta \tau \right)\right) e^{-i\pi f\ell\left( v_c^{-1}+v_{\pi}^{-1}\right)}
$$
are through S-parameters, while
$$
S_{41}^o = -2i \frac{\sqrt{p}}{1+p} \sin\left(\pi f\Delta \tau \right) e^{-i \pi f \ell\left( v_c^{-1}+v_{\pi}^{-1}\right)},
$$
$$
S_{23}^o = -2i \frac{\sqrt{p}}{1+p} \sin\left(\pi f\Delta \tau \right) e^{-i\pi f \ell\left( v_c^{-1}+v_{\pi}^{-1}\right)},
$$
are forward coupling S-parameters (see Appendix B).
Here, $f$ is the frequency and $\Delta \tau = \ell \left( v_c^{-1} - v_{\pi}^{-1} \right)$ represents the time delay difference between the $c$- and $\pi$- mode propagation in a coupled transmission line of length $\ell$. $\Delta \tau$ is zero for uncoupled transmission lines, and it becomes nonzero when forward coupling is present. $v_c$ and $v_{\pi}$ are propagation velocities for $c$- and $\pi$- modes respectively. The parameter $p = -R_c/R_{\pi}$ characterizes the asymmetry of the coupled transmission line. It equals one for a symmetric transmission line and deviates from one for an asymmetric transmission line, indicating the presence of intra-pair skew. The quantities $R_{c(\pi)}$ is defined as
\begin{align}\label{eq:RcRpi}
	 R_{c(\pi)} \! = \! (v_{c(\pi)}^{-2} \! - \! L_{11} C_{11} \! - \! L_{12} C_{21})/({L_{11} C_{12}+L_{12} C_{22}})\,.
\end{align}
The magnitudes of differential, common, and mixed-mode insertion losses can be calculated from (\ref{S_param_00}) (for further details on the derivations, see Appendix~\ref{App_B}). The asymmetry (skew) impact on the differential and common-mode insertion losses are equal and, when expanded in terms of the small parameter $\delta t_s=1-\sqrt{p}$ - which corresponds to a small asymmetry (skew) in the coupled transmission line-are given by
\begin{align}\label{eq:sdd21_0}
|S_{dd21}&|^2 =|S_{cc21}|^2\approx\cos^2{\left(\pi f t_l\right)}\nonumber \\
&-0.5\sin{\left(2\pi f\Delta \tau\right)}\sin{\left(2\pi f t_l\right)}\delta t_s\nonumber \\
&+\left[0.125\cos\left(2\pi f\left(\Delta \tau-t_l\right)\right)\right.\\
&\!+\!0.375\cos\!\left(2\pi \!f\left(\Delta \tau \!+\! t_l\right)\right) \left. \!- 0.5\cos{\left(2\pi f t_l\right)}\right] \delta t_s^2\,. \nonumber
\end{align}
The mixed-mode insertion loss (mode conversion), similarly expanded in terms of the small parameter $\delta t_s$, is given by
\begin{align}\label{eq:sdc21_0}
|S_{cd21}&|^2 \approx \sin^2{\left(\pi f t_l\right)}+0.5\sin{\left(2\pi f\Delta \tau\right)}\sin{\left(2\pi f t_l\right)}\delta t_s\nonumber \\
&-\left[0.125\cos\left(2\pi f\left(\Delta \tau-t_l\right)\right) \right. \\
&\!+\!0.375\cos\!\left(2\pi \!f\!\left(\Delta \tau \!+\! t_l\right)\right) \left. \!- 0.5\cos{\left(2\pi f t_l\right)}\right]\delta t_s^2. \nonumber
\end{align}
The magnitude of single-ended through S-parameters expanded in terms of the small parameter $\delta t_s$ can be written as follows:
\begin{align}\label{eq:s12_0}
|S_{21}|^2 = |S_{43}|^2 \approx\cos^2\left(\pi f \Delta \tau\right)
+ \sin^2\left(\pi f \Delta \tau\right) \delta t_s^2\,.
\end{align}
The forward coupling S-parameters are given by
\begin{align}\label{eq:s14_0}
|S_{41}|^2 =|S_{23}|^2 \approx\sin^2\left(\pi f \Delta \tau\right)  
- \sin^2\left(\pi f \Delta \tau\right) \delta t_s^2\,.
\end{align} 
The intra-pair skew can be derived (see Appendix~\ref{App_C}) for both signal direction left to right and right to left signal propagations (Fig.~ \ref{FIG:one} b).
The intra-pair phase skew at differential port 2 is defined as 
\begin{align}{\label{eq_skew1}}
	t_{\rm{skew,21}}=t_{1,2}-t_{2,2}\,,
\end{align}
where
\[
t_{1,2}={{\rm{phase}}}(S_{sd21})/(2\pi f) \hspace{1.5mm}\text{and}\hspace{1.5mm} t_{2,2}={{\rm{phase}}}(S_{sd41})/(2\pi f) \,,
\]
are time delays corresponding to the propagation of the signal from mixed-mode port 1 to the single-ended port 2 and port 4, respectively. $S_{sd21}=1/\sqrt{2}(S_{21}-S_{23})$ and $S_{sd41}=1/\sqrt{2}(S_{43}-S_{41})$ are S-parameters from mixed-mode port 1 to single-ended ports 2 and 4, respectively.
Similarly, the intra-pair skew at differential port 1 would be
\begin{align}{\label{eq_skew2}}
	t_{\rm{skew,12}} &= {{\rm{phase}}}(S_{sd12})/(2\pi f) \nonumber\\
	& - {{\rm{phase}}}(S_{sd32})/(2\pi f) \,.
\end{align}
Using (\ref{S_param_00}) intra-pair skew is calculated as follows
\begin{align}\label{eq:skew12_0}
t_{\mathrm{skew},12}\approx t_l + \Delta \tau \mathrm{sinc}\left(2\pi f \Delta \tau\right)\delta t_s\,,
\end{align}
and 
\begin{align}\label{eq:skew21_0}
t_{\mathrm{skew},21}\approx\cos\left(2\pi f \Delta \tau\right) t_l + \Delta \tau \mathrm{sinc}\left(2\pi f \Delta \tau\right)\delta t_s\,.
\end{align}
$\delta t_s$ can be estimated in terms of $c$- and $\pi$- mode phase velocities as follows (see Appendix~\ref{App_C}).
\begin{align*}
	\delta t_s \approx \frac{\delta c'+\delta c''}{v_c^{-2}-v_\pi^{-2}},
\end{align*}
where $\delta c'$ and $\delta c''$ (defined in Appendix~\ref{App_C}) creating the asymmetry between P and N lines.

Thus, the intra-pair skews from (\ref{eq:skew12_0}) and (\ref{eq:skew21_0}) can be expressed as follows:
\begin{align}\label{eq:skew_g1}
t_{\mathrm{skew},12}&\approx t_l + v_s^{-1} \ell \mathrm{sinc}\left(2\pi f \Delta \tau\right)\,,
\end{align}
and
\begin{align}\label{eq:skew_g2}
	t_{\mathrm{skew},21}&\approx\cos\left(2\pi f \Delta \tau\right) t_l
	+ v_s^{-1} \ell \mathrm{sinc}\left(2\pi f \Delta \tau\right)\,,
\end{align}
where $v_s=(v_c + v_\pi)/((\delta c'+\delta c'') v_c v_\pi )$.

\section{Special Cases and Behavior} 

In this section, we will consider special cases of transmission lines with both zero and non-zero forward coupling, analyzing scenarios with and without additional frequency independent time delay intra-pair skew $t_l$. We will examine the single-ended and balanced-mode S-parameters, as well as the intra-pair skews, in each of these cases.
\subsection{Zero Forward Coupling}
We start with considering case with zero forwarded coupling ($\Delta \tau =0$). This is an example of striplines in PCBs.

In these limits, regardless of the presence of additional intra-pair skew $t_l$, from (\ref{S_param_00}) the magnitudes of the single-ended S-parameters are given by:
\begin{align}\label{S_param_0}
	\left[ {\begin{array}{cc}
			|S_{21}|  & |S_{41}|  \\
			|S_{23}| & |S_{43}| \\
	\end{array} } \right]=
	\left[ {\begin{array}{cc}
			1  & 0 \\
			0 &1\\
	\end{array} } \right]\,.
\end{align}
The balanced modes as a function of additional intra-pair skew can be obtained from 
(\ref{eq:sdd21_0}) and (\ref{eq:sdc21_0}).
The differential and common mode insertion losses takes form:
\begin{align}\label{eq:sdd21_c0}
	|S_{dd21}|^2& =|S_{cc21}|^2\approx\cos^2{\left(\pi f t_l\right)}\,,
\end{align}
while differential to common mode insertion loss is
\begin{align}\label{eq:scd21_c0}
|S_{cd21}|^2 &\approx \sin^2{\left(\pi f t_l\right)}\,.
\end{align}
The intra-pair skew when forward coupling ($\Delta \tau =0$) is zero becomes symmetric and can be calculated from (\ref{eq:skew12_0}) and (\ref{eq:skew21_0}) as follows:
\begin{align}\label{eq:skew_c0}
	t_{\mathrm{skew},21}=t_{\mathrm{skew},12}=t_l\,.
\end{align}

 \begin{figure}[!t]
	\includegraphics[width=3.5in]{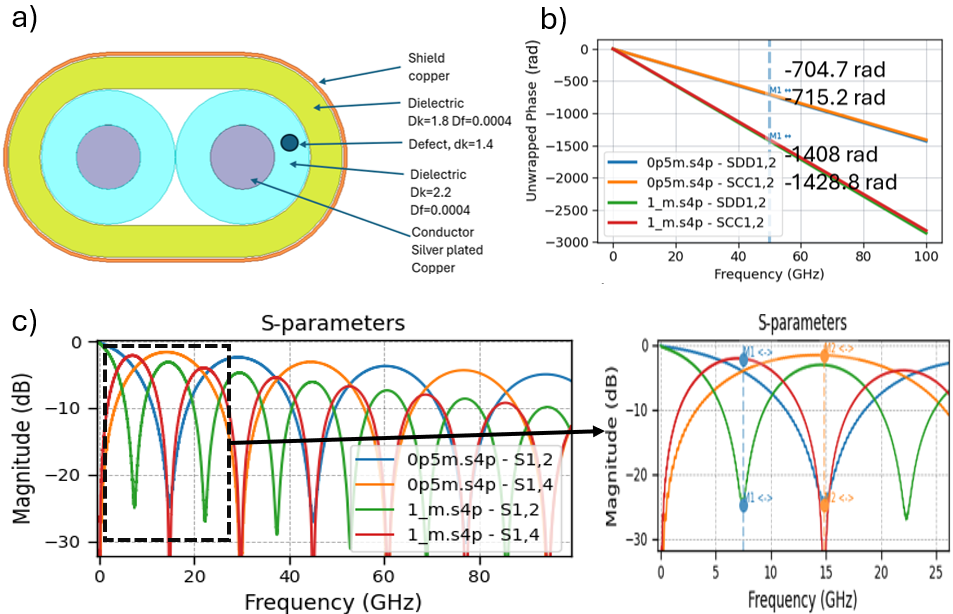}
	\caption{(a) Schematic of the cross section of dual-extruded twinax cable with 26 AWG conductors and differing dielectric regions used to model asymmetry.
		(b) Simulated single-ended through and forward-coupling S-parameters for 0.5m and 1m dual-extruded twinax cables, showing the periodic resonances caused by even and odd mode propagation delay differences.
		(c) Unwrapped phase plots for common and differential modes, used to extract $\Delta \tau$ for resonance frequency calculations.}
	\label{non_sym_sg}
\end{figure}

\subsection{Non-Zero Forward Coupling}
In this subsection, we consider the case of non-zero forward coupling ($\Delta \tau \ne 0$). This situation is typical in microstrips on PCBs, twinax cables, and certain high-speed connectors.

To validate the analytical solutions, we examine a 3D-modeled, dual-extruded twinax cable with 26 AWG wires and lengths of 0.5m and 1m in HFSS simulations. Intra-pair skew is introduced by modeling asymmetry via localized dielectric defects, where the defect dielectric constant is dk = 1.4 compared to the cable dielectric constant of dk = 2.2 (see Fig.~\ref{non_sym_sg} a).

\subsubsection{Single-ended S-parameters}

The magnitudes of through single-ended S-parameters are given by (\ref{eq:s12_0}) and  (\ref{eq:s14_0}). The additional frequency independent intra-pair skew $t_l$ does not affect the magnitude of the single-ended S-parameters. For asymmetric coupled lines, the effect of intra-pair skew is very small, since  $\delta t_s$ is small.

The through (\ref{eq:s12_0}) and forward coupling (\ref{eq:s14_0}) S-parameters oscillate with frequency, with a periodicity defined by the propagation time delay difference between the $c$-mode and $\pi$-mode, given by $2 / \Delta \tau$. For small asymmetry (intra-pair skew), $\Delta \tau$ can be approximated as the propagation time delay difference between the even and odd modes.

From the S-parameters, we can calculate the unwrapped phase to determine the propagation time delay difference between the even and odd modes as:
\begin{align}\label{eq:dTau}
\Delta\tau(f) = \frac{|{\rm{phase}}(S_{dd21}(f))-{\rm{phase}}(S_{cc21}(f))|}{2\pi f}\,.
\end{align}
Correspondingly, the resonance frequencies where the single-ended through S-parameter  (\ref{eq:s12_0}) reaches a minimum and forward coupling (\ref{eq:s14_0}) reaches a maximum are given by:
\begin{align}\label{eq:fn}
f_n=1/(2\Delta \tau)+ 1/(\Delta \tau)(n-1)\,,
\end{align}
where $n$ is the resonance order.

For our simulated 0.5m twinax cable, using (\ref{eq:dTau}) and the unwrapped phase values from Fig.~\ref{non_sym_sg} (b), we obtain $\Delta \tau_{0p5m}\approx 33.4$ps which, using of (\ref{eq:fn}), results in the first resonance at $f_1\approx 15$GHz. For the 1m twinax cable  $\Delta \tau_{1m}\approx 66.2$ps corresponding to $f_1\approx 7.6$GHz.
These resonance frequencies match well with the behavior observed in the single-ended through and forward-coupled S-parameters (see Fig.~\ref{non_sym_sg} c).

\subsubsection{Balanced Mode S-parameters when $t_l = 0$}
Let us start with the case where the additional intra-pair (frequency independent time delay) skew is zero ($t_l = 0$). 

\begin{figure}[!t]
	\includegraphics[width=3.5in]{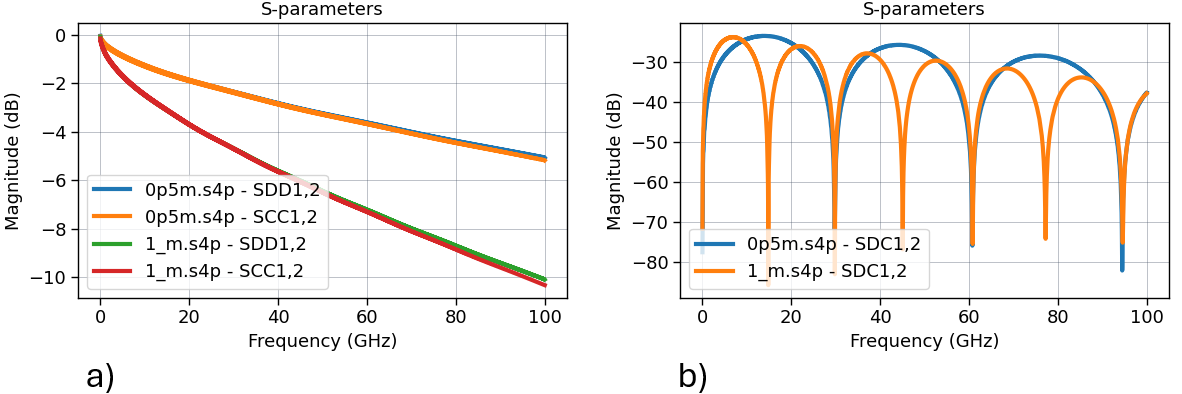}
	\caption{(a) Simulated differential- and common- mode insertion losses for 0.5m and 1m asymmetrical twinax cables, illustrating the minimal effect of intra-pair skew of cable itself.
		(b) Simulated differential-to-common mode conversion for 0.5m and 1m asymmetrical twinax cables, showing the asymmetry introduces minor mode conversion.}
	\label{non_sym_bm_tl0}
\end{figure}

In this situation, from (\ref{eq:sdd21_0}), both differential-mode and common-mode insertion losses are minimally affected by intra-pair skew. The influence appears only as very small oscillations, which are practically invisible, since the additional terms are on the order of ($\delta t_s$), a small quantity. These minor oscillations are superimposed on the zero-skew ($t_l=0$) case for both differential and common-mode insertion losses (see Fig.~\ref{non_sym_bm_tl0} a). The differential-to-common mode conversion, as given by (\ref{eq:sdc21_0}), is very small, on the order of ($\delta t_s$) which is consistent with the simulation results shown in Fig.~\ref{non_sym_bm_tl0} (b).
\subsubsection{Intra-pair skew when $t_l = 0$}
The intra-pair skew of the coupled transmission line (e.g. twinax cables) itself, from (\ref{eq:skew_g1}) and (\ref{eq:skew_g2}), is given by:
\begin{align}\label{eq:skew_bm_tl0}
	t_{\mathrm{skew},12}=t_{\mathrm{skew},21}\approx  v_s^{-1} \ell \mathrm{sinc}\left(2\pi f \Delta \tau\right)=t_s\mathrm{sinc}\left(2\pi f \Delta \tau\right),
\end{align}
where $t_s$ skew denotes oscillation amplitude in strong coupled transmission line.

\begin{figure}[!t]
	\centering
	\includegraphics[width=3in]{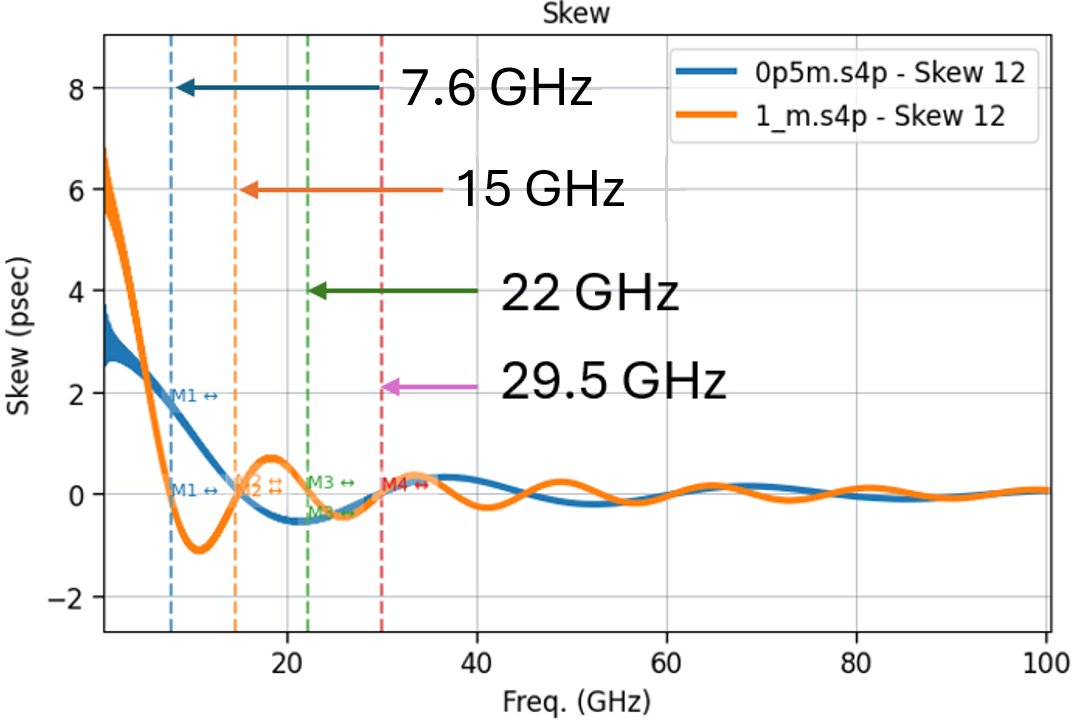}
	\caption{Intra-pair skew versus frequency for twinax cables of 0.5m and 1m length, showing the damped oscillatory behavior and zero-skew points predicted by (\ref{eq:skew_bm_tl0}) and (\ref{eq:skew_zero}).}
	\label{non_sym_sk_tl0}
\end{figure}

As shown in (\ref{eq:skew_bm_tl0}), this expression is symmetric, meaning that the intra-pair skew for signal propagation from left to right is identical to that for signal propagation from right to left. From this equation, it can be seen that in asymmetric coupled transmission lines, the intra-pair skew exhibits a damped oscillatory behavior as a function of frequency, attenuating at higher frequencies due to its dependence on the sinc function. Physically, this happens because in systems with strong forward coupling, any asymmetry between the P and N conductors does not solely affect one line; rather, it induces an additional oscillatory phase shift on both conductors. This supplementary phase shift is a oscillatory function of frequency, which implies that the resulting time delay  (defined as the phase divided by angular frequency) becomes frequency dependent. Consequently, as frequency increases, the additional time delay on each conductor decreases. Thus, intra-pair skew vanishes at higher frequencies (see Fig.~\ref{non_sym_sk_tl0}).
The intra-pair skew becomes zero at specific frequencies given by:
\begin{align}\label{eq:skew_zero}
f_0=\frac{n}{2\Delta \tau}\,,
\end{align}
where $n$ is the order of the zero corresponding to the intra-pair skew.

\subsubsection{Balanced Mode S-parameters when $t_l \ne 0$}
Next, we consider the case when the additional intra-pair skew $t_l$ is not zero. This means the channel consists of a constant (frequency-independent, flat) skew $t_l$ cascaded with an asymmetrically coupled transmission line (Fig.~\ref{FIG:one} a).

From (\ref{eq:sdd21_0}), we see that the extra insertion loss in both the differential and common modes due to $t_l$ is dominated by the constant intra-pair skew component, corresponding to the cosine-squared term in (\ref{eq:sdd21_0}). 

\begin{figure}[!t]
	\includegraphics[width=3.5in]{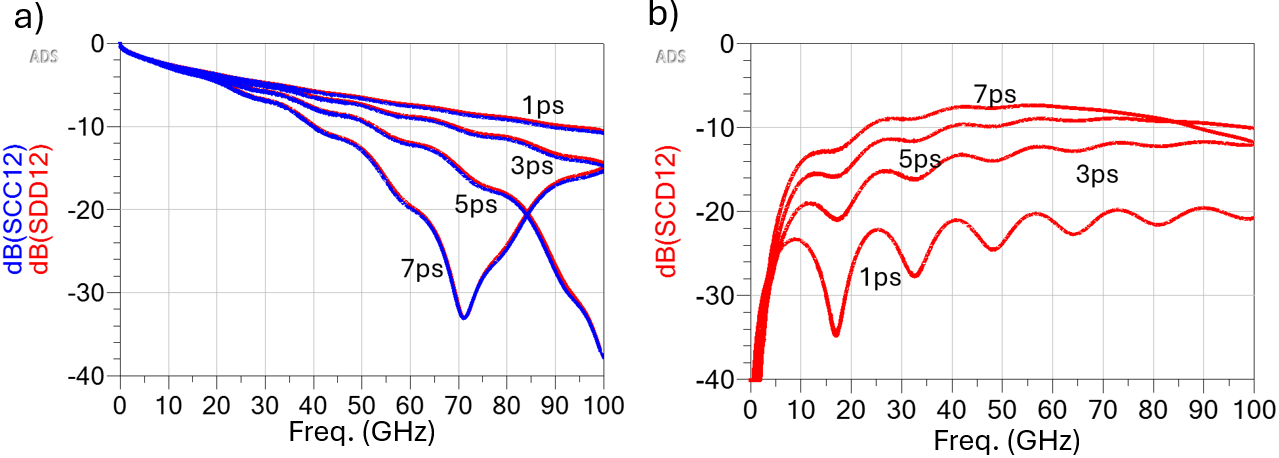}
	\caption{(a) Differential- and common- mode insertion losses for cascaded channels combining a frequency-independent $t_l$ skew for a 1m asymmetrical twinax cable. Simulations are shown for $t_l$  values from 1ps to 7ps in 2ps increments, illustrating the increasing visibility of oscillatory contributions as $t_l$  grows. 		
		(b) Differential-to-common mode conversion for cascaded channels combining a frequency-independent $t_l$  skew for a 1m asymmetrical twinax cable. Simulations are shown for $t_l$  values from 1ps to 7ps in 2ps increments, highlighting the stronger relative impact of oscillatory contributions when $t_l$  is small.} 
	\label{non_sym_bm_tl}
\end{figure}

To study this behavior in simulations, we sweep the constant intra-pair skew $t_l$ from 1ps to 7ps for a 1m asymmetrical twinax cable (Fig.~\ref{FIG:one} a). The resulting differential- and common- mode insertion losses are shown in Fig.~\ref{non_sym_bm_tl} (a). As seen, the behavior matches the analytical predictions (\ref{eq:sdd21_0}): the additional insertion loss is dominated by the cosine term with $t_l$, and the effect of the coupled asymmetrical cable’s  skew is minimal for small $t_l$. The oscillatory contribution becomes more visible as $t_l$ grows. 


Figure~\ref{non_sym_bm_tl} (b) illustrates the differential-to-common mode conversion for the same simulations presented in Fig.~\ref{FIG:one} (a). The behaviors observed in simulation match the analytical predictions: for larger $t_l$, mode conversion is dominated by first sine term of (\ref{eq:sdc21_0}), while for smaller $t_l$, the mode conversion is reduced in magnitude and exhibits oscillatory variations due to the intra-pair skew ($\delta t_s)$ of the asymmetrical twinax cable.

\subsubsection{Intra-pair skew when $t_l \ne 0$}
When the channel consists of a constant-$t_l$ intra-pair skew section cascaded with an asymmetrically coupled transmission line, the resulting intra-pair skews are not reciprocal, as shown in (\ref{eq:skew_g1}) and (\ref{eq:skew_g2}).

Specifically, when the constant-$t_l$ skew section is placed after the asymmetrical coupled line (in the direction of signal propagation, right-to-left), the skew observed for propagation through the coupled line and then the $t_l$ section can be calculated from (\ref{eq:skew_g1}). In this case, the damped oscillatory skew of the asymmetric coupled line is simply offset by an amount equal to $t_l$.

Conversely, when the signal propagates from the constant-$t_l$ skew section into the asymmetrical coupled line (left-to-right), the intra-pair skew is determined by (\ref{eq:skew_g2}). Here, the damped oscillatory skew shape of the asymmetric coupled line is modulated by an oscillatory cosine function with an amplitude determined by $t_l$, and with the same argument as the sinc function that describes the skew of the asymmetric coupled transmission line.

\begin{figure}[!t]
	\includegraphics[width=3.5in]{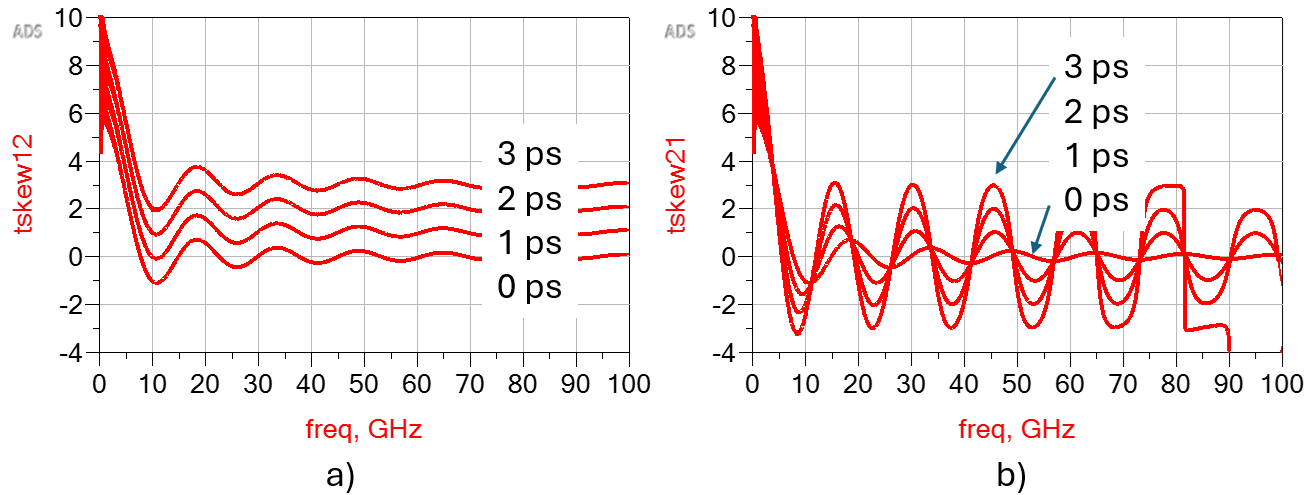}
	\caption{(a) Intra-pair skew as a function of frequency for signal propagation from the asymmetric twinax cable through the constant-$t_l$ skew section. The skew profile is offset by $t_l$, consistent with (\ref{eq:skew_g1}). Simulations are shown for $t_l$ values from 0ps to 3ps in 1ps increments.	
		(b) Intra-pair skew  as a function of frequency for signal propagation from the constant-$t_l$ skew section into the asymmetric twinax cable. The skew profile is modulated by an oscillatory function with amplitude $t_l$, as predicted by (\ref{eq:skew_g2}). Simulations are shown for $t_l$ values from 0ps to 3ps in 1ps increments.} 
	\label{non_sym_sk_tln}
\end{figure}

To validate the theoretical predictions, we examine the intra-pair skews from both ends of the simulated channel, similar to the approach used for Fig.~\ref{non_sym_bm_tl}, but with a different $t_l$ range—from 0ps to 3ps in step of 1ps. Figure~\ref{non_sym_sk_tln} shows the resulting skew profiles as functions of frequency. For signal propagation from the asymmetric twinax cable through the constant-$t_l$ skew section, the skew profile is offset by an amount equal to $t_l$, as predicted by (\ref{eq:skew_g1}). In the opposite direction—signal propagation from the constant-$t_l$ skew section into the asymmetric twinax cable—the skew profile of the cable is modulated by an additional oscillatory component with amplitude $t_l$, consistent with the analytical prediction in (\ref{eq:skew_g2}).

\section{Intra-pair Skew Propagation Graph (ISPG)} 
In this section, for complex channels consisting of multiple loosely and strongly coupled segments, we introduce the Intra-pair Skew Propagation Graph (ISPG). In Section II, we derived intra-pair skew equations for the case of cascading two channels: one with frequency-independent skew $t_l$ equivalent to a loosely coupled line with ($\Delta \tau \approx 0$), and a second asymmetric, strongly coupled ($\Delta \tau \ne 0$) channel, as given in (\ref{eq:skew_g1}) and (\ref{eq:skew_g2}).

To calculate the skew for channels comprising multiple sections of loosely and strongly coupled transmission lines, we used the Mathematica software package to solve configurations with three and four segments by cascading S-parameters (\ref{S_param_00}) and computing the cascaded channel intra-pair skews defined in (\ref{eq_skew1}) and (\ref{eq_skew2}). Based on the observed structure of the resulting expressions, we formulated a generalized approach in which the overall channel response is obtained through recursive cascading of section S-parameters. This formulation naturally extends to an arbitrary number of sections and to any ordering of loosely and strongly coupled lines, since each section is represented by a consistent multi-port network model and combined using standard network composition rules.

A complete end-to-end high-speed channel can thus be represented as a cascade of Loosely Coupled (LC) and Strongly Coupled (SC) segments. The total intra-pair skew of the full channel can be computed as follows:

Index the cascaded channel left→right as $i=1,\dots,N$, where $N$ is the number of segments to be cascaded.  Define the phase
$\Theta_m = \sum_{k=m}^N \theta_k\,,$
with $\Theta_{0}=0$ and
$$ \theta_k = \begin{cases}
	0, & \text{if stage }k\text{ is LC},\\
	2\pi f\Delta\tau_k, & \text{if stage }k\text{ is SC},
\end{cases} $$
with ${\dtau}_k=\tau_{\pi,k}-\tau_{c,k}$,  $\pi$- and $c$- modes propagation time difference of SC stage of channel. 
Let $t_{l,i}$ be the skew at LC stages, and $t_{s,i}=v_{s,i}^{-1}\ell_i$ is skew amplitude of SC stages. Then the skew contributions from both types of stages combine as
\begin{align}\label{eq:totalSkew}
		S=\sum_{i=1}^{N}&
		\big[
		\underbrace{t_{l,i}\cos(\Theta_{N-i+1})}_{\substack{\text{LC contributes}\\ \text{at current phase}}} \notag
		\\&+
		\underbrace{t_{s,i}\sinc\left(\pi f {\dtau}_i\right)\cos\left(\Theta_{N-i} + \pi f {\dtau}_i\right)}_{\substack{\text{SC contributes}\\ \text{mid-phase cosine}}}
		\big]\,.
\end{align}

\subsection{Counter Sweep Rule}
We introduce a set of simple sweep rules to evaluate (\ref{eq:totalSkew}). Assuming the signal propagates from left to right, the cumulative intra-pair skew is evaluated at the right-hand end of the channel. The steps are as follows:
\begin{enumerate} 
	\item Initialize phase $\ph=0$, and total skew $\;S=0$.
	\item Move right$\to$left:
	\begin{itemize}
		\item At an SC segment:
	\begin{itemize}	
		\item add contribution to skew:
		 $$S \gets S + t_{s}\sinc(\pi f\dtau)\cos(\ph+\pi f\dtau)\,.$$
		 	\item update phase with $\dtau$:
		 $$
		 \ph \gets \ph + 2\pi f\dtau\,.
		 $$
	\end{itemize}
	\item At an LC segment with skew $t_l$:
	\begin{itemize}		 	
		\item add contribution to skew: $$S \gets S + t_{l}\cos(\ph)\,.$$
		\item The phase $\ph$ stays same.
	\end{itemize}
		\end{itemize}
\end{enumerate}
\noindent At the far left, $S$ equals to \eqref{eq:totalSkew} exactly.

\subsection{Graph Visualization}
Next, we introduce graph-based rules to further simplify skew calculations, including the evaluation of the counter sweep rule. In this representation, the cascaded channel is modeled as a sequence of nodes, where each node corresponds to a distinct segment of the channel—either LC or SC—and the connecting edges represent skew propagation between adjacent segments (Fig.~\ref{g0}).

In this notation:

\begin{itemize} 
	\item LC - represented by a rectangle node labeled $t_l$, indicating a frequency-independent intra-pair skew.
	\item SC -  represented by a circular node labeled $\Delta\tau$ and $t_s$, indicating a strongly coupled segment with a propagation delay difference $\Delta\tau$ between $c$- and $\pi$- modes, and an associated skew term $t_s$.
\end{itemize}
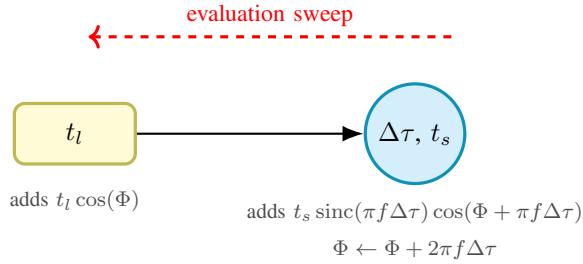
\begin{figure}[ht]
	\centering
	\begin{tikzpicture}[node distance=30mm]
		\node[lc] (l) {$t_l$};
		\node[sc, right=of l] (c) {$\Delta\tau$, $t_s$};
		
		\draw[arr] (l) -- (c);
		
		\draw[sweep] (c.north east) ++(0,0.8)
		-- node[above]{\small evaluation sweep} ++(-4.8,0);
		
		\node[annot, below=2mm of l] {adds $t_l\cos(\ph)$};
		\node[annot, below=1mm of c] {adds $t_s\sinc(\pi f\Delta\tau)\cos(\ph+\pi f\Delta\tau)$};
		\node[annot, below=6mm of c] {$\ph \gets \ph + 2\pi f\Delta\tau$};
	\end{tikzpicture}
	\caption{Example of the graph-based counter sweep rule for a cascaded LC–SC channel segments. 
		Here, $t_l$ and $t_s$ represent skew terms for loosely coupled and strongly coupled segments, respectively.
		The dashed red sweep arrow indicates the evaluation direction, while solid black arrow shows signal propagation.}
	\label{g0}
\end{figure}

\subsection{Worked Example}
We consider an example of cascaded loosely and strongly coupled channels to illustrate the application of the proposed graph-based rules. We demonstrate how the rules are applied step-by-step to calculate the intra-pair skew, and we compare the analytical results with simulation data to validate the method.

\begin{figure}[ht]
	\centering
	\begin{tikzpicture}[node distance=6mm]
		\node[lc] (p1) {$t_{l,1}$};
		\node[sc, right=of p1] (c1) {${\dtau}_A, t_{s,A}$};
		\node[lc, right=of c1] (p2) {$t_{l,2}$};
		\node[sc, right=of p2] (c2) {${\dtau}_B, t_{s,B}$};
		\draw[arr] (p1) -- (c1) -- (p2) -- (c2);
		\draw[sweep] (c2.north east) ++(0,0.8) -- node[above]{\small evaluation sweep} ++(-7.6,0);
	\end{tikzpicture}
		\caption{Graph representation of the cascaded channel LC($t_{l,1}$) $\to$ SC(${\dtau}_A, t_{s,A}$) $\to$ LC($t_{l,2}$) $\to$ SC(${\dtau}_B, t_{s,B}$). Rectangles (LC) denote loosely coupled segments with constant skew $t_l$, circles (SC) denote strongly coupled segments with delay difference $\Delta\tau$ and skew $t_s$. The dashed red sweep arrow indicates the evaluation direction, while solid black arrow shows signal propagation.}
	\label{g1}
\end{figure}
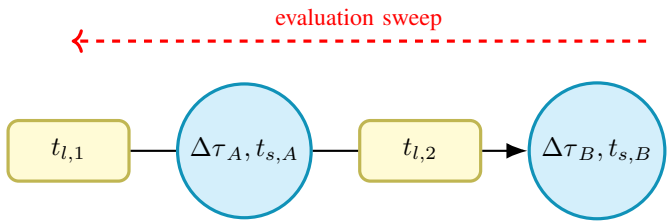

We begin by considering the channel configuration: LC($t_{l,1}$) $\to$ SC(${\dtau}_A, t_{s,A}$) $\to$ LC($t_{l,2}$) $\to$ SC(${\dtau}_B, t_{s,B}$). The corresponding graph representation is shown in Fig.~\ref{g1}.
Following to the counter sweep rules at each step we get
\begin{enumerate} 
	\item Start: phase $\ph=0$ and total skew $S=0$.
	\item SC(${\dtau}_B, t_{s,B}$): $$S \gets t_{s,B}\sinc(\pi f{\dtau}_B)\cos(\pi f{\dtau}_B)$$  and \\ $$\ph  \gets 2\pi f {\dtau}_B.$$
	\item LC($t_{l,2}$): $$\!\!\! \!\!\! S \! \gets \! t_{l,2} \cos(2\pi \!f\! {\dtau}_B)\!+\!t_{s,B}\sinc(\pi \!f\!{\dtau}_B)\!\cos(\pi \!f\!{\dtau}_B).$$
	\item SC(${\dtau}_A, t_{s,A}$): 
	\begin{align*}
	S& \gets t_{s,A}\sinc(\pi f{\dtau}_A)\cos(2\pi f {\dtau}_B+\pi f{\dtau}_A) \\
	&\!+ \!t_{l,2} \cos(2\pi \!f\! {\dtau}_B)\!+\!t_{s,B}\sinc(\pi \!f\!{\dtau}_B)\cos(\pi \!f\!{\dtau}_B)
	\end{align*}
	and
	 $$\ph \gets 2\pi f ({\dtau}_A+{\dtau}_B).$$
	\item LC($t_{l,1}$):
	\begin{align}\label{eq:wrk1}
		S& \gets t_{l,1} \cos\big(2\pi f({\dtau}_A+{\dtau}_B)\big)\nonumber \\ 
		&+t_{s,A}\sinc(\pi f{\dtau}_A)\cos(2\pi f {\dtau}_B+\pi f{\dtau}_A) \nonumber \\
		&+ t_{l,2} \cos(2\pi f {\dtau}_B) \nonumber \\
		&+t_{s,B}\sinc(\pi f{\dtau}_B)\cos(\pi f{\dtau}_B)\,.
	\end{align}
\end{enumerate}

\begin{figure}[!t]
	\includegraphics[width=3.5in]{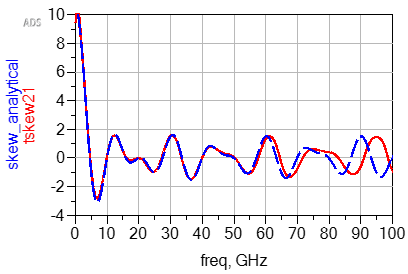}
	\caption{Intra-pair skew as a function of frequency for the channel configurations described in Section IV. Blue lines represent the analytically calculated skew, while red lines show the results from cascaded S-parameter simulations.} 
	\label{example1}
\end{figure}

Equation~(\ref{eq:wrk1}) represents the end-to-end intra-pair skew for the channel configuration under consideration. To validate the analytical result, we compare it with simulation data. Specifically, we consider a loosely coupled channel with skew $t_{l,1} = 0.5$ps connected to a 0.5m twinax cable characterized by $\Delta\tau_A = 33.4$ps and $t_{s,A} = 3$ps (low-frequency skew value for the 0.5m twinax cable from Fig.~\ref{non_sym_sk_tl0}). This is followed by another loosely coupled channel with skew $t_{l,2} = 1$ps, and finally connected to a 1m twinax cable with $\Delta\tau_B = 66.2$ps and $t_{s,B} = 6$ps (low-frequency skew value for the 1m twinax cable from Fig.~\ref{non_sym_sk_tl0}). Note that the values of $\Delta \tau$ are calculated in Section III using the approximated time-delay difference between the even- and odd- mode propagation delays. Figure~\ref{example1} shows the correlation between analytically calculated skew and simulations obtained by cascading the S-parameters of each channel section. 
As observed, the correlation is very strong, with only slight deviations at frequencies beyond 70GHz. These discrepancies arise because $\Delta \tau$ for the $c$- and $\pi$- mode propagation difference is approximated using the even- and odd- mode delay difference. This small approximation error becomes noticeable only at very high frequencies, manifesting as a slight horizontal shift in the skew profile (in frequency) without affecting the amplitude levels. Therefore, the method still predicts skew magnitudes accurately even at very high frequencies.

\section{Correlations to measurements} 
\begin{figure}[!t]
	\includegraphics[width=3.5in]{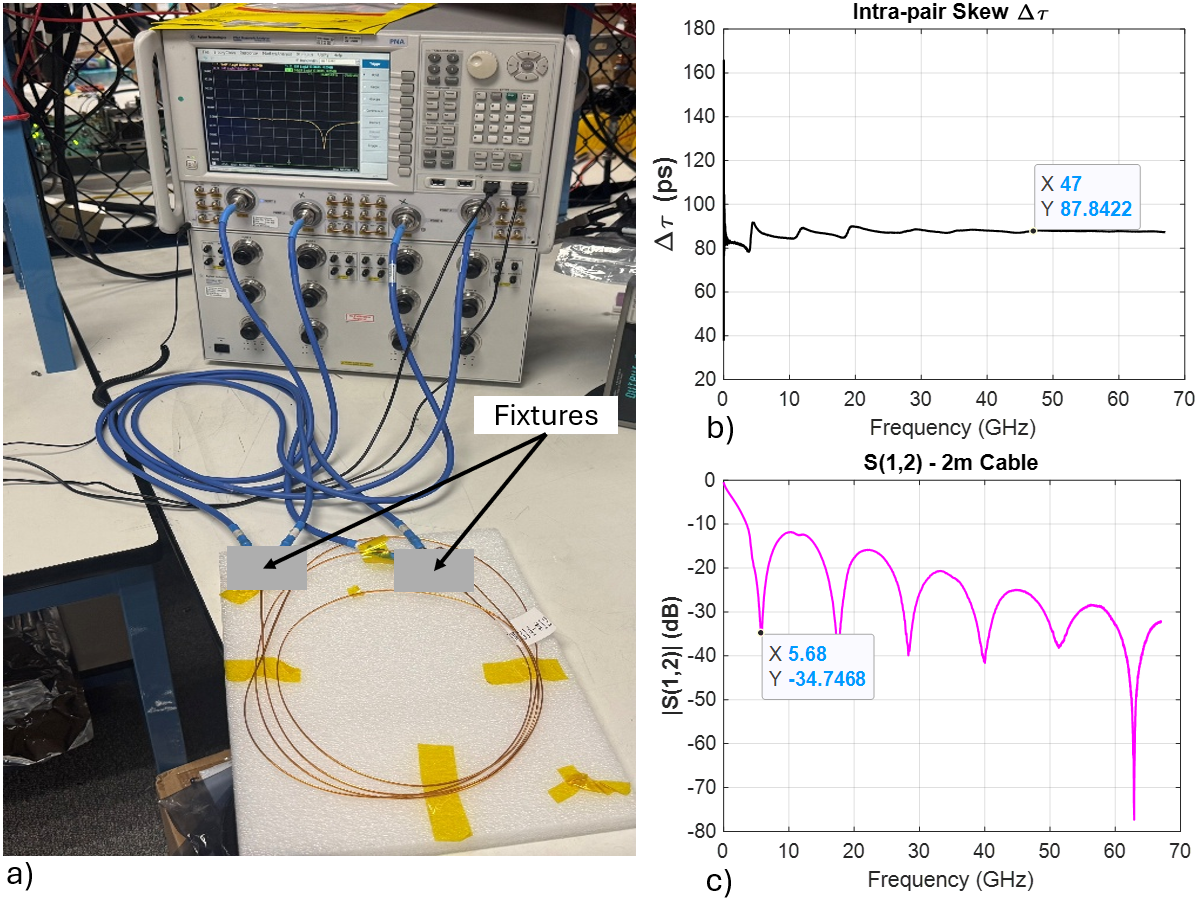}
	\caption{a) The skew measurements setup. b) The propagation time delay difference $\Delta \tau$  between the even- and odd- modes as function of frequency. c) Single-ended through S-parameters.} 
	\label{Fig:setup_deltaTau_S21}
\end{figure}

\begin{figure}[!t]
	\centering
	\includegraphics[width=3in]{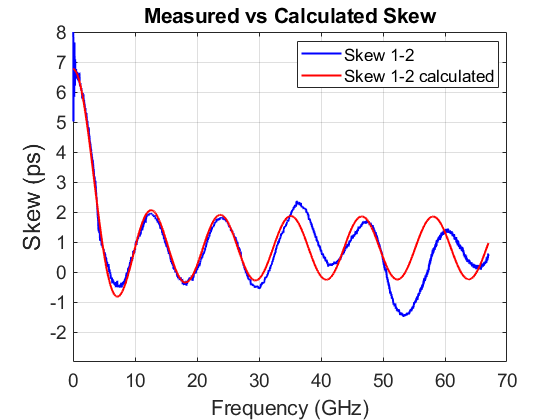}
	\caption{A comparison of the measured and estimated skew of bulk twinax cable.} 
	\label{Fig:Fig_meas_skew_1}
\end{figure}

To validate the skew estimation framework developed in Section IV, a series of measurements were performed using a 2m bulk twinax cable assembly. The measurement setup, depicted in Fig.~\ref{Fig:setup_deltaTau_S21} (a), consists of the bulk twinax cable connected to a Vector Network Analyzer (VNA) through fixtures with SMA cable connections at both ends. The measured topology of the 2m bulk twinax cable exhibits a cascaded structure with three distinct coupling regions: a loosely coupled section introduced by the input fixture, a strongly coupled region corresponding to the bulk twinax cable, and a loosely coupled section arising from the output fixture. Using this topology and the framework developed in Section IV, the estimated skew of the 2m bulk twinax cable is deduced as (\ref{eq:wrk2}). The propagation time delay difference  (let's say $\Delta\tau\approx 88$ps at 47GHz)  between the even- and odd modes- for this topology was computed using (\ref{eq:dTau}) and is shown in Fig.~\ref{Fig:setup_deltaTau_S21} (b). The resonance frequency calculated from (\ref{eq:fn}) using $\Delta\tau$ also closely matches the first resonance frequency of the single-ended insertion loss of the bulk twinax cable including fixtures at both ends, as shown in Fig.~\ref{Fig:setup_deltaTau_S21} (c).
Figure~\ref{Fig:Fig_meas_skew_1} presents a comparison between the skew extracted from S-parameter measurements (using (\ref{eq_skew1})) and the theoretical skew derived from (\ref{eq:totalSkew}). The estimation parameters are defined as $t_{l,1}=0.8$ps, $t_s=5$ps, $t_{l,2}=1$ps and $\Delta\tau=88$ps. As shown in the Fig.~\ref{Fig:Fig_meas_skew_1}, the theoretical model exhibits excellent agreement with the measured data, validating the accuracy of the proposed estimation methodology
\begin{align}\label{eq:wrk2}
	\text{Skew}_{calc}& = t_{l,1} +t_s \sinc(\pi f \Delta\tau)\cos(\pi f \Delta\tau) \nonumber\\
	&+ t_{l,2} \cos(2 \pi f \Delta\tau).
\end{align}

 \begin{figure}[!t]
	\includegraphics[width=3.5in]{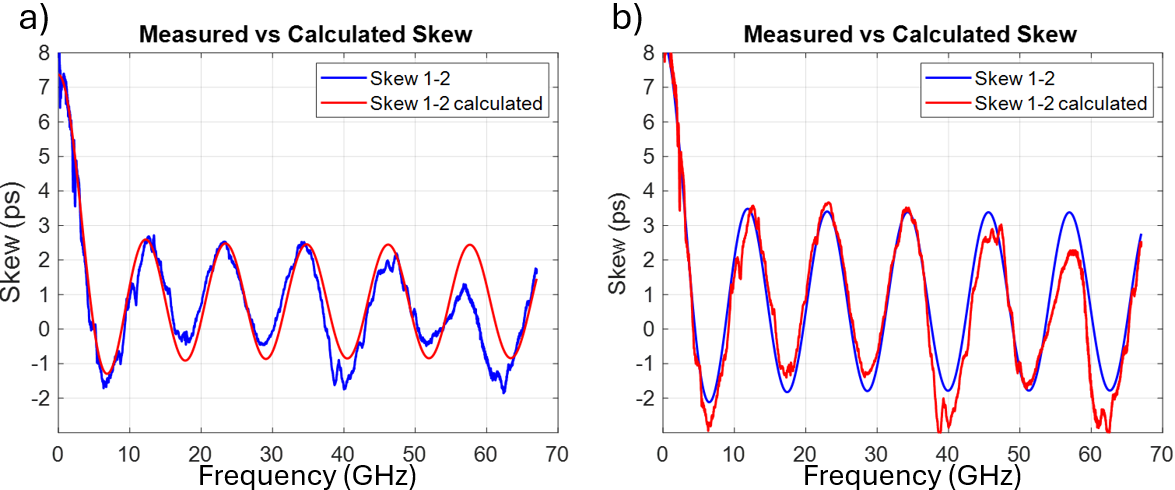}
	\caption{Comparison of measured and estimated skew for a bulk twinax cable cascaded with a differential PCB stripline channel exhibiting intra-pair skews of a) 1ps and b) 2ps.} 
	\label{Fig:Fig_meas_skew_2}
\end{figure}

For the second and third configurations, a differential PCB stripline channel exhibiting intra-pair skews of 1ps and 2ps, respectively, was connected through length matched coaxial cables at ends of the fixtures attached to the 2m bulk twinax cable. Because the stripline acts as a loosely coupled transmission line, it introduces an additional term into the cascaded structure. This extended topology is captured by the modified skew estimation formulation presented in (\ref{eq:wrk3}), where $t_{StripLine}$ is set to 1ps and 2ps for the respective configurations. The measured skew values were compared against the corresponding theoretical estimates, as shown in Figs.~\ref{Fig:Fig_meas_skew_2} (a) and (b). In both instances, the measured and estimated skew show excellent agreement, confirming that the extended model accurately captures the contribution of the differential stripline channel. The consistency of this agreement across all three configurations—baseline, +1ps, and +2ps—demonstrates the robustness of the proposed skew estimation methodology
\begin{align}\label{eq:wrk3}
	\text{Skew}_{calc} &= t_{l,1} + t_s \sinc(\pi f \Delta\tau)\cos(\pi f \Delta\tau) \nonumber \\
	&+ t_{l,2} \cos(2 \pi f \Delta\tau) + t_{StripLine} \cos(\pi f \Delta\tau).
\end{align}

\section{Conclusions} 
The rapid advancement of high-speed digital systems necessitates accurate and efficient modeling of differential interconnects, where intra-pair skew has emerged as a critical performance limiter. This paper addressed the existing gap in analytical frameworks by presenting a comprehensive derivation of intra-pair skew equations for generic asymmetric coupled transmission lines, explicitly integrating skew into the formulation of S-parameters.

A key contribution of this work is the development of closed-form expressions that accurately predict intra-pair skew and its impact on balanced-mode S-parameters for both loosely and strongly coupled transmission lines. We demonstrated that for channels with non-zero forward coupling, the intra-pair skew exhibits a damped oscillatory behavior as a function of frequency, attenuating at higher frequencies. Furthermore, the skew becomes non-reciprocal when a constant frequency-independent skew $t_l$ is cascaded with an asymmetric coupled lines. Specifically, the direction of signal propagation significantly alters the skew profile, manifesting either as a simple offset or as an amplitude modulation of the twinax cable's intrinsic skew.

To facilitate the analysis of complex interconnects, we introduced the Intra-pair Skew Propagation Graph (ISPG), a novel graph-based methodology for efficiently calculating end-to-end intra-pair skew in cascaded channels composed of multiple loosely and strongly coupled segments. This approach provides a systematic "counter sweep rule" for evaluating cumulative skew, which was rigorously validated against detailed S-parameter simulations for various channel configurations. The excellent correlation between analytical predictions and simulation results, even at high frequencies (up to 70GHz and beyond), confirms the robustness and accuracy of the proposed framework.

This analytical tool offers designers a powerful and computationally efficient alternative to time-consuming 3D electromagnetic simulations. By providing clear insight into how intra-pair skew modifies transmission and coupling characteristics, this work enables more effective prediction of skew-induced impairments, ultimately facilitating the design of more robust and higher-performance high-speed interconnects.

Finally, the proposed framework was successfully validated through empirical measurements of a 2m bulk twinax cable assembly. The close agreement between the analytical model and the measured skew across all three configurations—baseline, +1ps, and +2ps—demonstrates the robustness of the methodology. By effectively accounting for both strongly and loosely coupled transmission line sections, this framework provides a reliable and versatile tool for predicting skew in complex, cascaded interconnect structures.

\appendices
\section{Derivation of the Phase Velocities}\label{App_A}

To find the general solution for coupled system (\ref*{eq:teleg}), we seek fundamental solutions in the following form:
\[
\begin{bmatrix} V_1(z) \\ V_2(z) \end{bmatrix}=\begin{bmatrix} n_1 \\ n_2 \end{bmatrix} e^{\gamma z},
\]
where $\gamma$ is the eigenvalue of system (\ref*{eq:teleg}) satisfying the following characterization equation
\begin{align}\label{eq:eign}
	\gamma^4+\omega^2 (P_{11}+P_{22} ) \gamma^2+\omega^4 P_{11} P_{22}- \omega^4 P_{12} P_{21}=0,
\end{align}
where
\begin{align}\label{eq:Pmat}
	\left[ {\begin{array}{cc}
			\! \! \! P_{11} & \! P_{12} \! \! \! \\
			\! \! \! P_{21} & \! P_{22} \! \! \! \\
	\end{array} } \right]\! \! =\! \! 
\left[ {\begin{array}{cc}
	\! \! \! L_{11} C_{11}\! +\! L_{12} C_{21} &\! L_{11} C_{12}\! +\! L_{12} C_{22} \! \! \! \\
		\! \! \! L_{21} C_{11}\! +\! L_{22} C_{21} & \! L_{21} C_{12}\! +\! L_{22} C_{22} \! \! \! \\
\end{array} } \right].
\end{align}

The $c$- and $\pi$- mode phase velocities can be found from the following equation using the complex propagation constants $\gamma_{c(\pi)}$, found from the (\ref{eq:eign})
\[
v_{c(\pi)} \! =\! \frac{\omega}{\gamma_{c(\pi)}} \! = \! \sqrt{\frac{2}{L_{11} C_{11} \! + \! L_{12} C_{21} \! + \! L_{21} C_{12} \! + \! L_{22} C_{22} \mp u}},
\]
where
\[
u = \left((L_{11} C_{11}+L_{12} C_{21} - L_{21} C_{12}-L_{22} C_{22} )^2 \right.
\]
\[
\left. +4(L_{11} C_{12}+L_{12} C_{22})(L_{21} C_{11}+L_{22} C_{21})\right)^{1/2}.	
\]

\section{Derivation of S-parameters}\label{App_B}

The forward-traveling normalized voltage waves \cite{Mongia_1999} on the two lines, described by (\ref{eq:teleg}), can be expressed as a linear combination of $c$- and $\pi$- mode voltage waves as follows:
\begin{align*}
	V_1 (z)&=\frac{A_1}{\sqrt{Z_{c1}}} e^{\gamma_c z}+\frac{A_2}{\sqrt{Z_{\pi1}}} e^{\gamma_\pi z}, \\
	V_2 (z)&=\frac{A_1 R_c}{\sqrt{Z_{c2}}} e^{\gamma_c z}+\frac{A_2 R_\pi}{\sqrt{Z_{\pi2}}} e^{\gamma_\pi z}.
\end{align*}

Here, the coefficients $A_1$ and $A_2$  are determined by boundary conditions, $R_c$ and $R_{\pi}$ defined as (\ref{eq:RcRpi}) and
\begin{align*}
	Z_{c1}\! =\! \frac{ i\omega\! \left( L_{12}L_{21} \! - \! L_{11} L_{22}\right) }{\gamma_c \left( L_{12} R_c - L_{22}\right) }, \,
	Z_{c2}\! =\! \frac{i\omega R_c \! \left( L_{12}L_{21} \! - \! L_{11} L_{22} \right)}{\gamma_c\left( L_{21} - L_{11} R_c\right) }, \\
	Z_{\pi1}\! =\! \frac{ i\omega\! \left( L_{12}L_{21} \! - \! L_{11} L_{22}\right) }{\gamma_{\pi} \left( L_{12} R_{\pi} - L_{22}\right) }, \,
	Z_{\pi2}\! =\! \frac{i\omega R_{\pi} \! \left( L_{12}L_{21} \! - \! L_{11} L_{22} \right)}{\gamma_{\pi}\left( L_{21} - L_{11} R_{\pi}\right) }.
\end{align*}

After applying boundary conditions $V_1 (0)=1, V_2 (0)=0$, we get
\begin{align*}
	A_1 = \frac{\sqrt{Z_{c1}}}{1+p}, \quad A_2 = \frac{p\sqrt{Z_{\pi1}}}{1+p},
\end{align*}
where $p=-R_c/R_{\pi}$.

So, we have the following solutions:
\begin{align*}
	&V_1 (z)=\frac{1}{1+p} e^{-i\beta_c z}+\frac{p}{1+p} e^{-i\beta_\pi z} \\
	&=\! e^{-i \frac{\beta_c + \beta_\pi}{2} z} \left[\cos\! \left(\frac{\beta_c\! -\! \beta_\pi}{2} z\right)\! -\! i \frac{1-p}{1+p} \sin\! \left(\frac{\beta_c\! -\! \beta_\pi}{2} z\right)\right],
\end{align*}
\begin{align*}
	V_2 (z)&=\frac{\sqrt{Z_{c1}} R_c}{\sqrt{Z_{c2}} (1+p)} e^{-\beta_c z}-\frac{\sqrt{Z_{\pi1}} R_\pi}{\sqrt{Z_{\pi2}} (1+p)} \frac{R_c}{R_\pi} e^{-\beta_\pi z} \\
	&=-2i \frac{\sqrt{p}}{1+p} e^{-i \frac{\beta_c + \beta_\pi}{2} z} \sin\left(\frac{\beta_c-\beta_\pi}{2} z\right),
\end{align*}
where $\beta_{c(\pi)}=-i\gamma_{c(\pi)} = 2\pi f v_{c(\pi)}^{-1}$.

Thus, the $S^o_{21}$ and $S^o_{41}$ can be found as voltage at the end of the transmission lines:
\begin{align*}
	S^o_{21}&=V_1 (\ell)\\
	&=e^{-i \pi f\left(v_{c}^{-1} + v_{\pi}^{-1} \right)\ell} \left[\cos\left( \pi f \Delta \tau\right)-i \frac{1-p}{1+p} \sin\left( \pi f \Delta \tau\right)\right], \\
	S^o_{41}&=V_2 (\ell)=-2i \frac{\sqrt{p}}{1+p} e^{-i \pi f \left( v_{c}^{-1} + v_{\pi}^{-1}\right)  \ell} \sin\left( \pi f \Delta \tau\right),
\end{align*}
where $\Delta \tau = \left(v_{c}^{-1}-v_{\pi}^{-1}\right) \ell$.

Now, let us find $S^o_{23}$ and $S^o_{43}$. Applying the following boundary conditions $V_1 (0)=0, V_2 (0)=1$, we arrive at:
\begin{align*}
	A_1=\frac{\sqrt{Z_{c2}} \sqrt{Z_{c1} Z_{\pi2}}}{R_c \sqrt{Z_{c1} Z_{\pi2}}-R_\pi \sqrt{Z_{\pi1} Z_{c2}}}, \\
	A_2=-\frac{\sqrt{Z_{c2}} \sqrt{Z_{\pi2}} \sqrt{Z_{\pi1}}}{R_c \sqrt{Z_{c1} Z_{\pi2}}-R_\pi \sqrt{Z_{\pi1} Z_{c2}}},
\end{align*}
and consequently, after simple transformation we get the following solutions:
\begin{align*}
	V_1 (z)=-\frac{2i\sqrt{p}}{1+p} e^{-i \frac{\beta_c + \beta_\pi}{2} z} \sin\left(\frac{\beta_c-\beta_\pi}{2} z\right),
\end{align*}
\begin{align*}
	V_2 (z)\!=\!e^{-i \frac{\beta_c + \beta_\pi}{2} z}\! \left[\! \cos\!\left(\frac{\beta_c\!-\!\beta_\pi}{2} z\right)\! + \!i \frac{1\! -\! p}{1\! +\! p}\! \sin\!\left(\frac{\beta_c\!-\!\beta_\pi}{2} z\right)\! \right].
\end{align*}
Thus,
\begin{align*}
	S^o_{23}&=V_1 (\ell)=-2i \frac{\sqrt{p}}{1+p} e^{-i \pi f\left(v_{c}^{-1} + v_{\pi}^{-1} \right) \ell} \sin\left(\pi f \Delta \tau\right),\\
	S^o_{43}&=V_2 (\ell) \\
	&=\! e^{-i \pi f\left(v_{c}^{-1} + v_{\pi}^{-1} \right) \ell} \! \left[\cos\left(\pi f \Delta \tau\right)\! +\! i \frac{1\! -\! p}{1\! +\! p} \sin\left(\pi f \Delta \tau\right)\! \right]. \\
\end{align*}
\vspace{-15mm}
\section{Estimation of the Intra-pair Skew Term and Phases}\label{App_C}

In the first section, we used term $1-\sqrt{p}$ in (\ref{eq:sdd21_0})-(\ref{eq:s14_0}), (\ref{eq:skew12_0}), and (\ref{eq:skew21_0}) which is responsible for the intra-pair skew occurrence in strongly coupled transmission lines.

Let $\delta c'$ and $\delta c''$ be model asymmetry between P and N conductors in (\ref{eq:Pmat}) thus creates
intra-pair skew, and they are much smaller than $P_{22}$ and $P_{21}$ respectively
\begin{align*}
	P =	\left[ {\begin{array}{cc}
			P_{22}+\delta c' & P_{21} + \delta c'' \\
			P_{21} & P_{22} \\
	\end{array} } \right],
\end{align*}

Recalling expressions for $R_c$ and $R_{\pi}$ we get
\begin{equation}\label{eq:1-p}
	\begin{aligned}
		1-\sqrt{p} &= 1-\sqrt{-\frac{R_c}{R_\pi}} \\
		&= -\frac{\delta c'+\delta c''}{2|P_{21}|} +O\left(\left( \delta c' \right)^2+\left( \delta c'' \right)^2\right).
	\end{aligned}
\end{equation}

On the other hand,
\begin{equation}\label{eq:v_cp}
	\begin{aligned}
		v_\pi^{-2}&-v_c^{-2}=\frac{(-i\gamma_{\pi} )^2}{\omega^2} - \frac{(-i\gamma_c )^2}{\omega^2}\\
		&=2|P_{21}| + \mathrm{sign}(P_{21}) \delta c'' + O\left(\left( \delta c' \right)^2+\left( \delta c'' \right)^2\right).
	\end{aligned}
\end{equation}

Applying (\ref{eq:1-p}) and (\ref{eq:v_cp}), finally we arrive at the following estimation
\begin{align}\label{eq:1-p_final}
	1-\sqrt{p}&\approx \frac{\delta c'+\delta c''}{v_c^{-2}-v_\pi^{-2}}.
\end{align}

Now let us estimate phases for some parameters from the Section II. Recalling expressions for $S_{21}$, $S_{23}$, $S_{43}$, $S_{41}$, and expanding in series near $1-\sqrt{p}$, it is not difficult to get
\begin{align*}
	S_{21} &\! - \! S_{23} \! = \! e^{-i\pi f \ell (v_c^{-1} + v_{\pi}^{-1})} [\cos(\pi f\Delta\tau) \! - \! 2\pi f \sin(\pi f\Delta\tau)x t_l \\
	&\! + \! i\left(2\pi f \cos(\pi f\Delta\tau)t_l \! + \! (x-1)\sin(\pi f\Delta\tau\right)\! +\! O(x^2 +t_l^2)],
\end{align*}
and
\begin{align*}
	S_{43} &- S_{41} = e^{-i\pi f \ell (v_c^{-1} + v_{\pi}^{-1})} [\cos(\pi f\Delta\tau) + 2\pi f \sin(\pi f\Delta\tau) t_l \\
	&- i(x+1)\sin(\pi f\Delta\tau) + O(x^2 +t_l^2)],
\end{align*}
where $x=1-\sqrt{p}$.

Thus, applying estimation (\ref{eq:1-p_final}), the phases can be estimated as follows:
\begin{align*}
	\mathrm{phase}(S_{sd21}) &= \mathrm{phase}(S_{21} - S_{23}) \approx -\pi f \ell (v_c^{-1} + v_{\pi}^{-1})\\
	& -\pi f\Delta\tau + 0.5\sin(2\pi f\Delta\tau)\frac{\delta c'+\delta c''}{v_c^{-2}+v_\pi^{-2}} \\
	& +2\pi f\cos^2(\pi f\Delta\tau)t_l,
\end{align*}
\begin{align*}
	\mathrm{phase}(S_{sd41}) &= \mathrm{phase}(S_{43} - S_{41}) \approx  -\pi f \ell (v_c^{-1} + v_{\pi}^{-1})\\
	& -\pi f\Delta\tau - 0.5\sin(2\pi f\Delta\tau)\frac{\delta c'+\delta c''}{v_c^{-2}+v_\pi^{-2}} \\
	& +2\pi f\sin^2(\pi f\Delta\tau)t_l.
\end{align*}

Similarly,
\begin{align*}
	\mathrm{phase}(S_{sd12}) &\! = \! \mathrm{phase}(S_{12} - S_{14}) \\
	&\approx -\pi f (\ell v_c^{-1} + \ell v_{\pi}^{-1} - 2t_l)\\
	& - \pi f\Delta\tau + 0.5\sin(2\pi f\Delta\tau)\frac{\delta c'+\delta c''}{v_c^{-2}+v_\pi^{-2}}\,,
\end{align*}
\begin{align*}
	\mathrm{phase}(S_{sd14}) &= \mathrm{phase}(S_{34} - S_{32}) \approx -\pi f \ell (v_c^{-1} + v_{\pi}^{-1})\\
	&-\pi f\Delta\tau - 0.5\sin(2\pi f\Delta\tau)\frac{\delta c'+\delta c''}{v_c^{-2}+v_\pi^{-2}}\,.
\end{align*}

\bibliographystyle{IEEEtran}
\bibliography{b}

\end{document}